\documentclass[runningheads]{llncs}
\usepackage{multirow}
\usepackage{bbding}
\usepackage[T1]{fontenc}
\usepackage{graphicx}
\begin{document}
\fontsize{9pt}{11pt}\selectfont
%
\title{Joint Training or Not: An Exploration of Pre-trained Speech Models in Audio-Visual Speaker Diarization }
%
%
\author{Huan Zhao\inst{1} \and
Li Zhang\inst{1} \and
Yue Li\inst{1} \and 
Yannan Wang \inst{2} \and Hongji Wang \inst{2} \and Wei Rao \inst{2} \and Qing Wang \inst{1} \and Lei Xie \inst{1}
}
\authorrunning{Huan Zhao et al.}
%
\institute{Audio, Speech and Language Processing Group (ASLP@NPU), School of Computer Science, \\
Northwestern Polytechnical University (NPU), Xi'an, China \and
Tencent Corporation, Shenzhen, China}
%
\maketitle              
\begin{abstract}
The scarcity of labeled audio-visual datasets is a constraint for training superior audio-visual speaker diarization systems. To improve the performance of audio-visual speaker diarization, we leverage pre-trained supervised and self-supervised speech models for audio-visual speaker diarization. Specifically, we adopt supervised~(ResNet and ECAPA-TDNN) and self-supervised pre-trained models~(WavLM and HuBERT) as the speaker and audio embedding extractors in an end-to-end audio-visual speaker diarization~(AVSD) system. Then we explore the effectiveness of different frameworks, including Transformer, Conformer, and cross-attention mechanism, in the audio-visual decoder. To mitigate the degradation of performance caused by separate training, we jointly train the audio encoder, speaker encoder, and audio-visual decoder in the AVSD system. Experiments on the MISP dataset demonstrate that the proposed method achieves superior performance and obtained third place in MISP Challenge 2022.

\end{abstract}
\section {Introduction}
Speaker diarization~(SD) is the task of determining “who spoke when?” in an audio or video recording~\cite{DBLP:journals/taslp/MiroBEFFV12}. Its application is indispensable in multimedia information retrieval, speaker turn analysis, and multi-speaker speech recognition. With the emergence of wide and complex application scenarios of SD, single-modal~(audio- or visual-based) SD encounters performance bottleneck~\cite{watanabe2020chime,chen2022audio,zhou2022audio}. Specifically, speech is intrinsically uncertain because it includes multiple speech signals produced by multi-speakers, interfered by echoes, other audio sources, and ambient noise. Likewise, identifying speakers from a single visual modal is highly complex, and it is restricted to detecting lip or facial movements from frontal close-up images of people. In more general scenarios, such as informal social gatherings, people may not always face the cameras, which makes lip reading a difficult task~\cite{gebru2017audio}. To enhance the effectiveness of speaker diarization, it is crucial to utilize both visual and audio information. 

Recently, several works have begun to leverage both visual and audio cues for speaker diarization. Sharma et. al~\cite{sharma2022using} performed face clustering on the active speaker faces and showed superior speaker diarization performance compared to the state-of-the-art audio-based diarization methods. Sarafianos et al.~\cite{sarafianos2016audio} applied a semi-supervised version of Fisher Linear Discriminant analysis, both in the audio and the video signals to form a complete multimodal speaker diarization system.  In addition to introducing the facial cue as the visual modal, several methods were proposed to leverage audio and lip cues for diarization motivated by the synergy between utterances and lip movements~\cite{chung2019said}. Wuerkaixi et al.~\cite{wuerkaixi2022dyvise} proposed a dynamic vision-guided speaker embedding~(DyViSE) method in a multi-stage system to deal with the challenge in noisy or overlapped speech and off-screen speakers where visual features are missing. Tao et al.~\cite{tao2021someone} proposed an audio-visual cross-attention mechanism for inter-modality interaction between audio and visual features to capture long-term speaking evidence. He et al.~\cite{he2022} proposed a novel end-to-end audio-visual speaker diarization~(AVSD) method consisting of a lip encoder, a speaker encoder, an audio encoder, and an audio-visual speaker decoder. The end-to-end training strategy is similar to~\cite{tao2021someone,wuerkaixi2022dyvise}. Although AVSD obtains a comparable improvement compared with single-modal speaker diarization, the separate optimization of speaker encoder and audio-visual speaker decoder is hard to get global optima.
Furthermore, despite the demonstrated effectiveness of pre-trained speech models in various downstream speech tasks~\cite{DBLP:journals/taslp/HsuBTLSM21,DBLP:journals/jstsp/ChenWCWLCLKYXWZ22}, there has been limited research investigating the impact of self-supervised pre-training models on AVSD.

To fill this gap, we employ supervised pre-trained speaker models~(ResNet~\cite{DBLP:conf/cvpr/HeZRS16} and ECAPA-TDNN~\cite{DBLP:conf/interspeech/DesplanquesTD20}) and self-supervised models~(HuBERT~\cite{DBLP:journals/taslp/HsuBTLSM21} and WavLM~\cite{DBLP:journals/jstsp/ChenWCWLCLKYXWZ22}) as the speaker and audio embedding extractors in an 
 end-to-end AVSD system. Subsequently, we evaluate the effectiveness of different frameworks, such as Transformer~\cite{DBLP:conf/nips/VaswaniSPUJGKP17}, Conformer~\cite{gulati2020conformer}, and cross-attention mechanism~\cite{tao2021someone,wuerkaixi2022dyvise}, in the audio-visual decoder module of the AVSD system. Furthermore, to mitigate the decline in diarization performance that may arise from separate training, we jointly train the pre-trained audio encoder and speaker encoder as well as the audio-visual decoder modules in the AVSD system. Finally, the experimental results on the MISP dataset~\cite{2022misptask2} demonstrate joint training of self-supervised per-trained audio encoder, supervised pre-trained speaker encoder,
and Transformer-based audio-visual decoder obtains considerable improvements compared with the results of AVSD~\cite{he2022}.

\section {Method}

\begin{figure*}[bpth] 
\centering
\centerline{\includegraphics[width=\linewidth]{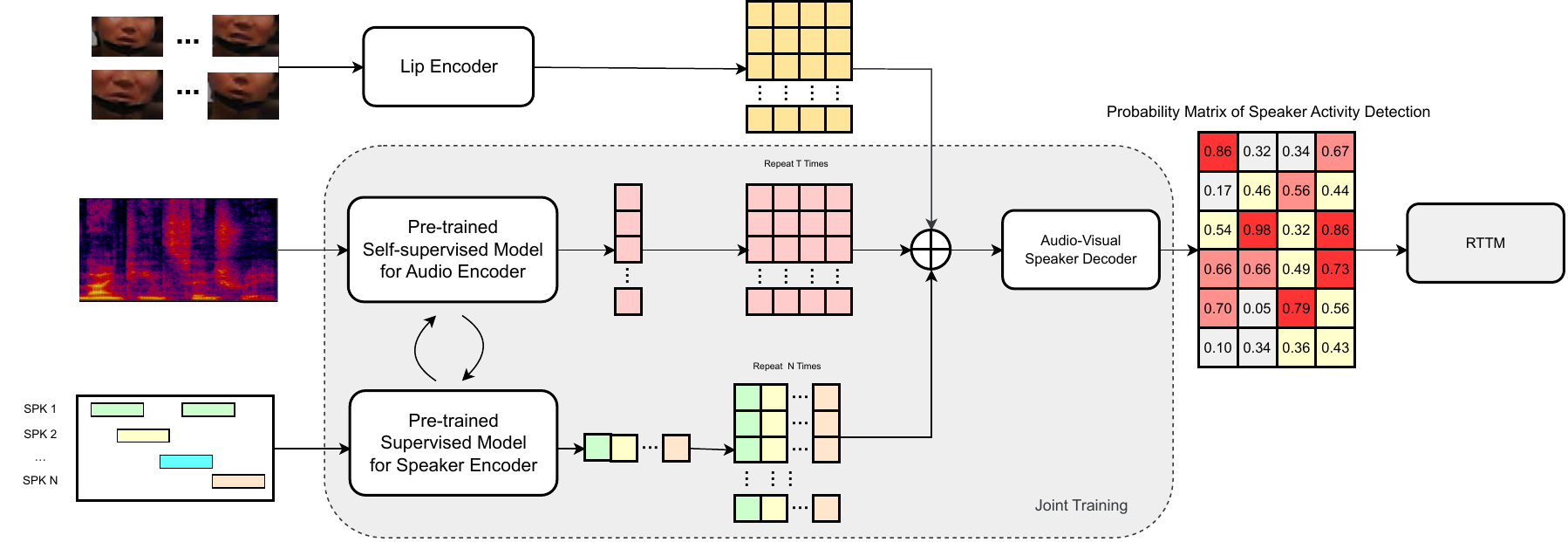}}
\caption{Overview of the improved AVSD framework.Parts in oval boxes with gray backgrounds require joint training. Arrows represent the same model being capable of serving as both the audio encoder and the speaker encoder simultaneously.
}
\label{fig:overview}
\vspace{-1em}%
\end{figure*}
\subsection {Overview}

An overview of the improved AVSD framework is illustrated in Figure~\ref{fig:overview}, which consists of a lip encoder, an audio encoder, a speaker encoder, and an audio-visual speaker decoder. Different from the AVSD framework in~\cite{he2022}, we adopt the pretrained self-supervised speech models~(WavLM and Hubert) and pre-trained supervised speaker models~(ResNet and ECAPA-TDNN) as our audio encoder and speaker encoder. Meanwhile, we jointly train the pre-trained speaker and audio encoders with the audio-visual speaker decoder to alleviate performance loss caused by separate training. Furthermore, we introduce different neural network frameworks (Transformer~\cite{DBLP:conf/nips/VaswaniSPUJGKP17}, Conformer~\cite{gulati2020conformer}, cross-attention~\cite{tao2021someone}) in the audio-visual decoder to effectively fusion of audio and visual representations and decode out the probability of speaker activity detection. Finally, the probability matrix is converted to Rich Transcription Time Masked~(RTTM). 

\subsection {Lip Encoder}

The lip encoder is configured as~\cite{he2022}, consisting of ResNet18~\cite{DBLP:conf/cvpr/HeZRS16}, Multi-Scale Temporal Convolutional Networks~(MS-TCN) followed by three Conformer~\cite{gulati2020conformer} and a BLSTM~\cite{DBLP:journals/neco/HochreiterS97} layers. Firstly, we cut the lip regions of interest~(ROIs) from the ground truth detection results from~\cite{2022misptask2}. Assume that we have a sequence of lip ROIs $X_n = (X_{n}^1, X_{n}^2, \dots, X_{n}^T) \in \mathbb{R}^{W \times H \times T}$, where $n$ is the identity of speaker and $T$ is the frame length. Then the ROIs are fed into the lip encoder. Then we stack embeddings of each speaker to get visual embeddings $E_{V} \in \mathbb{R}^{T \times N \times D^V}$. Moreover, the visual embeddings $E_{V}$ are projected through a fully connected~(FC) layer to speech activity probabilities $S = (S_{1}, S_{2}, \dots, S_{T}) \in (0, 1)$. Therefore, the lipreading model can be considered as a visual speaker activity detection module, which has the capability to represent whether the state is speaking or not of a speaker. In the end, this module provides a visual representation of whether a speaker is speaking to the AVSD framework as the visual cues.

\subsection {Supervised Pre-trained Models for Audio/Speaker Encoder}

Supervised pre-trained speaker models trained with large-scale datasets have a super generalization ability~\cite{zhang2022npu}. In this paper, we use them as an audio encoder, the utterance-level pooling in the pre-trained model is replaced with frame-level pooling. Before the pooling layer, assume that we have the convolutional bottleneck feature map  $M \in \mathbb{R}^{T \times D'}$, where $T$ and $D'$ are the numbers of frames and dimensions. Then we first uniformly split the feature map into short segments with a length of 5 frames and a shift of 1 frame. The dimensions of segment feature map $M_i$ are $T' \times D'$, where $T'$ is the length of the segment. Then we perform the utterance-level pooling on each segmented feature map. The frame-level pooling layer aggregates the two-dimensional feature map as follows: 

\begin{equation}
\mu_i = \frac{1}{T'} \sum_{t=1}^{T'} M_{i,t} ,
\end{equation}

\begin{equation}
\sigma_i = \sqrt{\frac{1}{T'} \sum_{t=1}^{T'} (M_{i,t} - \mu_i)^2},
\end{equation}
where $\mu_i$ and $\sigma_i$ are the mean and standard deviation vector of the $i$-th segment. Then, the concatenation of $\mu_i$ and $\sigma_i$ represents $i$-th frame pooling result. Finally, we stack all frame pooling results as audio embeddings $E_{A} \in \mathbb{R}^{T \times D^A}$.

When the supervised pre-trained speaker models are used as a speaker encoder, the supervised models are applied to extract utterance-level speaker embeddings. In the training stage, we extract speaker embeddings $E_S \in \mathbb{R}^{D^A}$ with non-overlapped speech segments of each speaker in oracle labels. During the decoding stage, we compute speaker embeddings with the non-overlapped speech segments which are estimated by the result of visual speaker activity detection.

\subsection {Self-supervised Pre-trained Models for Audio/Speaker Encoder}

The pre-trained self-supervised models learned from large-scale datasets have shown super generalizability and discrimination, which shows consistent improvements across downstream speech tasks \cite{DBLP:conf/nips/BaevskiZMA20,DBLP:conf/naacl/DevlinCLT19}. Inspired by this, we investigate two classical self-supervised pre-trained models~(HuBERT~\cite{DBLP:journals/taslp/HsuBTLSM21} and WavLM~\cite{DBLP:journals/jstsp/ChenWCWLCLKYXWZ22}) in our AVSD system.



Similar to the utilization of supervised pre-trained speaker models as mentioned above, we conduct an investigation into the impacts of employing HuBERT and WavLM as the speaker encoder and the audio encoder within the end-to-end AVSD system. To harness the generalization and discriminative capabilities inherent in pre-trained models, we cross-validated the effect of self-supervised and supervised pre-trained speech models as speaker encoders and audio encoders in the AVSD system.  

\subsection {Audio-visual Speaker Decoder}

The audio-visual speaker decoder estimates the target-speaker voice activities by considering the correlations between each speaker. We explore Transformer~\cite{DBLP:conf/nips/VaswaniSPUJGKP17}, Conformer~\cite{gulati2020conformer} and cross-attention~\cite{tao2021someone} in the decoder module respectively. Given visual embeddings $E_V$, audio embeddings $E_A$, and speaker embeddings $E_S$, firstly audio embedding $E_A$ are repeated $N$ times and speaker embeddings are repeated $T$ times, where $N$ denotes the number of speakers and $T$ is the length of frames. Then, we can obtain fusion audio-visual embeddings $E_F$ by concatenating visual, audio, and speaker embeddings. For the Transformer, a speaker state detection module consisting of a two-layer Transformer encoder with six blocks is designed to extract the states of each speaker. Then, we apply another Transformer encoder with six blocks to estimate the correlations between each speaker. The utilization of Conformer is similar to that of Transformer. For the cross-attention decoder, we first perform cross-attention between lip and speaker embeddings extracted from the pre-trained speaker encoder. Then another cross-attention is further performed between the output of the first cross-attention and audio embeddings from the audio encoder. Finally, a linear layer is considered to obtain the outputs corresponding to the speech/non-speech probabilities for each of the N speakers respectively.

\section {EXPERIMENTAL SETUP}

\subsection {Dataset}

We conduct experiments on the Multimodal Information Based Speech Processing~(MISP) dataset~\cite{DBLP:conf/icassp/ChenZDLCWSSLYPG22}. MISP is a dataset for home TV scenarios, which provides over 100 hours of audio and video recordings of 2-6 people in a living room interacting with smart speakers/TVs while watching and chatting. The supervised pre-trained model is trained on VoxCeleb~\cite{DBLP:conf/interspeech/NagraniCZ17}\cite{DBLP:conf/interspeech/ChungNZ18} and CN-Celeb~\cite{DBLP:conf/icassp/FanKLLCCZZCW20}. Then fine-tuned on the MISP training dataset. The self-supervised pre-trained model is trained on LibriSpeech dataset~\cite{DBLP:conf/icassp/PanayotovCPK15}. Table \ref{tab:dataset} details the specific data for training each module.

\begin{table}[th]
\caption{The training data used for each module.}
\label{tab:dataset}
\centering
\scriptsize
\resizebox{\columnwidth}{!}{
\begin{tabular}{lc}
\toprule
\hline
\multicolumn{1}{c}{Module}    & Training data     \\ \hline
Lip encoder     & MISP                \\ 
Supervised pre-trained model & \begin{tabular}[c]{@{}l@{}}CN-CELEB, VoxCeleb, MISP \end{tabular}     \\
Self-supervised pre-trained model & LibriSpeech    \\ 
Audio-visual speaker decoder   & MISP      \\ 
\bottomrule
\end{tabular}}
\end{table}
\vspace{-13pt}
\subsection {Configurations}
In this study, we employ ResNet34~\cite{DBLP:conf/cvpr/HeZRS16} with channel configurations of {32, 64, 128, 256} and ECAPA-TDNN(1024)~\cite{DBLP:conf/interspeech/DesplanquesTD20} as supervised pre-trained models. The embedding dimensions for ResNet34 and ECAPA-TDNN are set at 128 and 192, respectively. The loss function is additive angular margin softmax~(AAM-softmax)~\cite{DBLP:conf/cvpr/DengGXZ19} with a margin of 0.2 and a scale of 32. The speaker embedding models are trained with 80-dimensional log Mel-filter bank features with 25ms window size and 10ms window shift. We replace the utterance-level pooling of ResNet34 and ECAPA-TDNN with frame-level pooling to extract frame-level features. In addition, HuBERT Base and WavLM Base are utilized as self-supervised pre-trained models.  The back-end fusion module consists of three layers of Transformer encoder. Each encoder layer has 6 blocks with the same config: 256-dim attentions with 2 heads and 1024-dim feed-forward layers. The Adam optimizer with a learning rate of 1e-4 is used for training. During training, the pre-trained model is frozen in the first 5 epochs and then unfrozen to joint training for another 10 epochs. 

\subsection{Comparison Methods}

We compare the improved AVSD with other competitive multi-modal speaker diarization methods. They are listed in the following:
\begin{itemize}
     \item AVSD~\cite{he2022}: The author introduces an end-to-end audio-visual speaker diarization method that employs CNN networks to extract audio embeddings and ivectors for speaker embeddings~(CNN+ivector).

     \item WHU~\cite{cheng2023whu}: The author expands upon the Sequence-to-Sequence Target-Speaker Voice Activity Detection framework to concurrently identify the voice activities of multiple speakers from audio-visual signals.

     \item SJTU~\cite{liu2023multi}: The author integrates Interchannel Phase Difference (IPD) to represent spatial characteristics and pre-trains a model based on ECAPA-TDNN to extract speaker embedding features.

    \item Cross-attention~\cite{tao2021someone}: This work introduces a cross-attention mechanism~\cite{tao2021someone} in the temporal dimension to effectively integrate audio-visual interactions. 
    \item Conformer~\cite{gulati2020conformer}: The authors present a method that combines convolutional neural networks (CNN) and Transformer to obtain both local and global information for sequences.
\end{itemize}

\subsection {Evaluation Metric}

We use Diarization Error Rate~(DER)~\cite{DBLP:conf/mlmi/IstrateFMBB05} as the evaluation metric, which computes the ratio of incorrectly attributed speaking segments or non-speech segments. DER consists of three components: False Alarm Rate (FA), Missing Speech Rate (MISS), and Speaker Error Rate (SPKERR). Lower values for all three metrics indicate superior performance of the speaker diarization system. 
\section {Results and Analysis}

We present the results of the proposed audio-visual speaker diarization model in Table~\ref{tab:results-supervised-self-supervised}. The lip encoder is not always involved in training. For comparison, we show CNN-AE~(joint) + ivector-SE~(fixed) as a baseline and the corresponding DER is reported as 13.09\%. When we use a pre-trained model, the effect improved by 9.82\% compared to AVSD. Furthermore, when involving the audio encoder in joint training, DER further decreased by  28.86\%. Training both audio and speaker encoders jointly leads to a 17.32\% relative reduction on DER. In addition, we also compare the proposed method with other audio-visual methods. As shown in Table~\ref{tab:results-supervised-self-supervised}, the effect of our method is better than other audio-visual methods except WHU's.

\vspace{-10pt}
\begin{table}[th]
\caption{The results~(\%) of joint training with supervised and self-supervised model in AVSD. The audio encoder is abbreviated AE and the speaker encoder is abbreviated SE. (* represents joint training in the corresponding module).}
\label{tab:results-supervised-self-supervised}
\centering
\scriptsize
\resizebox{0.8\textwidth}{!}{
\begin{tabular}{lcccc}
\toprule
\hline
\multicolumn{1}{c}{Method}          & MISS & FA   & SPKERR & DER  \\ \hline
CNN-AE~\&~ivector-SE~\cite{he2022}    & 4.01 & 5.86 & 3.22   & 13.09 \\
WHU~\cite{cheng2023whu}  & - & - & - & 8.82\\
SJTU~\cite{liu2023multi}  & 4.44 & 4.82 & 2.10 & 10.82 \\
HuBERT-AE~\&~ResNet-SE & 3.72 & 5.66 & 2.53   & 11.92 \\
HuBERT-AE$^*$~\&~ResNet-SE & \textbf{1.88} & \textbf{4.96} & \textbf{2.41}   & \textbf{9.25} \\ 
HuBERT-AE$^*$~\&~ResNet-SE$^*$ & 2.35  & 5.14  & 2.67   & 10.16 \\ \bottomrule 
\end{tabular}}
\end{table}
\vspace{-10pt}

We also investigate the impact of different decoders on AVSD. Table~\ref{tab:results-decoder} displays that the DER is 9.97\% when using BLSTM as the decoder. When replacing BLSTM with the Conformer, the DER is reduced to 9.61\%. Cross-attention method is also employed, leading to a further reduction in DER to 9.57\%. Additionally, the best result of 9.54\% is obtained when using Transformer as the decoder.

\begin{table}[th]
\caption{DER~(\%) comparison among different audio-visual decoders in AVSD. Both the audio encoder and the speaker are ResNet.}
\label{tab:results-decoder}
\centering
\scriptsize
\resizebox{0.8\textwidth}{!}{
\begin{tabular}{lcccc}
\toprule
\hline
\multicolumn{1}{c}{Method} & MISS & FA   & SPKERR & DER  \\ \hline
BLSTM                       & 2.19 & 5.49 & 2.22   & 9.97 \\
Conformer                  & 2.01 & 5.50 & 2.10   & 9.61 \\
Cross-Attention            & 1.35 & 6.26 & 1.95   & 9.57 \\
Transformer                & \textbf{1.36} & \textbf{6.23} & \textbf{1.92}  & \textbf{9.54} \\ 
\bottomrule
\end{tabular}}
\end{table}

Next, we apply ablation experiments to assess the combined effect of supervised and self-supervised pre-trained models. Table~\ref{tab:results-supervised} shows the experiments using only the supervised pre-trained model for training. Initially, after replacing the audio encoder with ResNet, DER decreased from 9.25\% to 9.54\%. Then, replacing the speaker encoder with ivectors led to a further decrease to 12.16\%. Finally, replacing ResNet with CNN resulted in a DER reduction to 13.09\%. We also performed a replacement with ECAPA-TDNN, which yielded results close to ResNet. The results indicate that the supervised model contributes more to the speaker encoder, which is related to the model's training objective.

\vspace{-10pt}
\begin{table}[th]
\caption{DERs~(\%) of only using supervised model in AVSD(* represents joint training the corresponding module).}
\label{tab:results-supervised}
\centering
\resizebox{0.8\textwidth}{!}{
\begin{tabular}{lcccc}
\toprule
\hline
\multicolumn{1}{c}{Method}             & FA   & MISS & SPKERR & DER   \\ \hline
HuBERT-AE$^*$~\&~ResNet-SE & \textbf{1.88} & \textbf{4.96} & \textbf{2.41}   & \textbf{9.25} \\
ResNet-AE$^*$~\&~ResNet-SE  & 1.36 & 6.32 & 1.92   & 9.54  \\
ECAPA-AE$^*$~\&~ECAPA-SE~    & 2.13 & 5.14 & 2.13  & 9.68\\
ResNet-AE$^*$~\&~ivector-SE~ & 1.54 & 8.04 & 2.58   & 12.16 \\
CNN-AE$^*$~\&~ivector-SE    & 4.01 & 5.86 & 3.22   & 13.09 \\

\bottomrule[1pt]
\end{tabular}}
\end{table}
\vspace{-10pt}

Table~\ref{tab:results-self-supervised} illustrates the experiments using only the self-supervised model. Firstly, after replacing the speaker encoder with HuBERT, the model's performance decreased from 9.25\% to 10.14\%. Then, replacing HuBERT with WavLM led to a further decrease in DER to 11.35\%. Regardless of which self-supervised pre-trained model is used, the results are better than the baseline.

\begin{table}[th]
\caption{DERs~(\%) of only using self-supervised model in AVSD(* represents joint training the corresponding module).}
\label{tab:results-self-supervised}
\centering
\resizebox{0.8\textwidth}{!}{
\begin{tabular}{lcccc}
\toprule
\hline
\multicolumn{1}{c}{Method}             & MISS & FA   & SPKERR & DER   \\ \toprule[1pt]
HuBERT-AE$^*$~\&~ResNet-SE & \textbf{1.88} & \textbf{4.96} & \textbf{2.41}   & \textbf{9.25} \\
HuBERT-AE$^*$~\&~HuBERT-SE  & 2.84 & 5.01 & 2.29   & 10.14 \\
WavLM-AE$^*$~\&~WavLM-SE    & 2.03 & 5.84 & 3.48   & 11.35 \\
CNN-AE$^*$~\&~ivector-SE    & 4.01 & 5.86 & 3.22   & 13.09 \\
\bottomrule[1pt]
\end{tabular}}
\end{table}

\begin{figure}[th] 
\captionsetup{font={footnotesize}} 
\centering
\centerline{\includegraphics[scale=0.9]{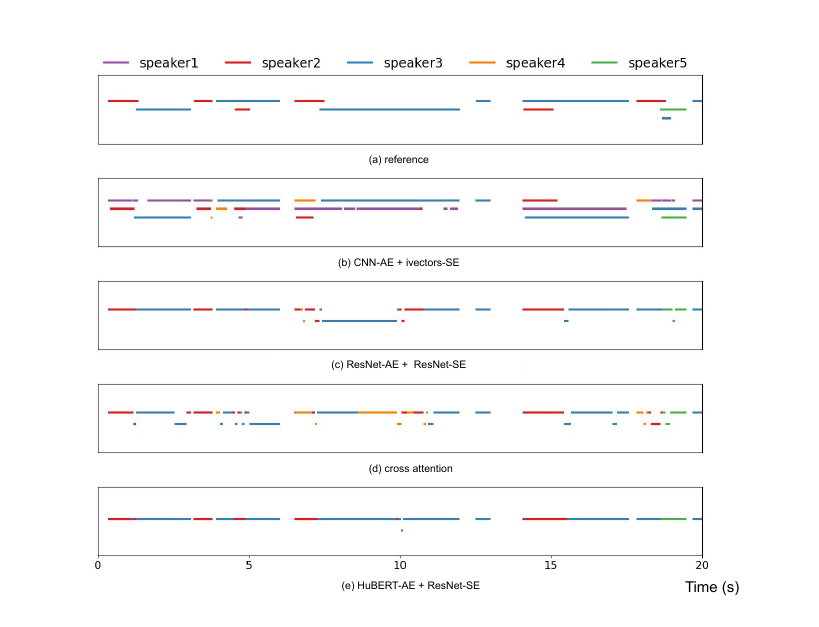}}
\caption{Visualization of the difference between the results of different methods and the reference results. 
}
\label{fig:visualization}
\end{figure}

Figure~\ref{fig:visualization} visualizes the RTTM on different methods. In comparison with the reference, it is observed that the baseline method exhibits numerous false alarms of other speakers. Then we replace the audio encoder and speaker encoder with ResNet, which can accurately predict the results. Furthermore, we perform a comparative analysis between the Transformer decoder and cross-attention methods. The latter demonstrated speaker errors around the 10th second. Moreover, the audio encoder is replaced with HuBERT, resulting in an improvement in detection.
\vspace{-10pt}
\section{Conclusion}

In this paper, we use the pre-trained supervised and self-supervised speech models as speaker encoders and audio encoders to improve the AVSD framework. To mitigate the performance decline caused by separate learning, we jointly train the speaker encoder, audio encoder, and audio-visual speaker decoder. In addition, we explore different audio-visual speaker decoders in the AVSD framework, which are Transformers, Conformers, and cross-attention respectively. By applying HuBERT as the audio encoder and ResNet as the speaker encoder, we achieve the best performance. Compared with the AVSD, the improved AVSD has a considerable reduction in DER on the MISP dataset.

%
%
%
\bibliographystyle{splncs04}
\bibliography{mybibliography}

\begin{thebibliography}{10}
\providecommand{\url}[1]{\texttt{#1}}
\providecommand{\urlprefix}{URL }
\providecommand{\doi}[1]{https://doi.org/#1}

\bibitem{DBLP:conf/nips/BaevskiZMA20}
Baevski, A., Zhou, Y., Mohamed, A., Auli, M.: wav2vec 2.0: {A} framework for self-supervised learning of speech representations. In: NeurIPS (2020)

\bibitem{2022misptask2}
Chen, H., Du, J., et~al., D.Y.: Audio-visual speech recognition in misp2021 challenge: Dataset release and deep analysis. In: Interspeech (2022)

\bibitem{chen2022audio}
Chen, H., Du, J., Dai, Y., Lee, C.H., Siniscalchi, S.M., Watanabe, S., Scharenborg, O., Chen, J., Yin, B.C., Pan, J.: Audio-visual speech recognition in misp2021 challenge: Dataset release and deep analysis. In: Proceedings of the Annual Conference of the International Speech Communication Association, INTERSPEECH. vol.~2022, pp. 1766--1770 (2022)

\bibitem{DBLP:conf/icassp/ChenZDLCWSSLYPG22}
Chen, H., Zhou, H., Du, J., Lee, C., Chen, J., Watanabe, S., Siniscalchi, S.M., Scharenborg, O., Liu, D., Yin, B., Pan, J., Gao, J., Liu, C.: The first multimodal information based speech processing (misp) challenge: Data, tasks, baselines and results. In: {ICASSP}. pp. 9266--9270. {IEEE} (2022)

\bibitem{DBLP:journals/jstsp/ChenWCWLCLKYXWZ22}
Chen, S., Wang, C., Chen, Z., Wu, Y., Liu, S., Chen, Z., Li, J., Kanda, N., Yoshioka, T., Xiao, X., Wu, J., Zhou, L., Ren, S., Qian, Y., Qian, Y., Wu, J., Zeng, M., Yu, X., Wei, F.: Wavlm: Large-scale self-supervised pre-training for full stack speech processing. {IEEE} J. Sel. Top. Signal Process.  \textbf{16}(6),  1505--1518 (2022)

\bibitem{cheng2023whu}
Cheng, M., Wang, H., Wang, Z., Fu, Q., Li, M.: The whu-alibaba audio-visual speaker diarization system for the misp 2022 challenge. In: ICASSP 2023-2023 IEEE International Conference on Acoustics, Speech and Signal Processing (ICASSP). pp.~1--2. IEEE (2023)

\bibitem{chung2019said}
Chung, J.S., Lee, B.J., Han, I.: Who said that?: Audio-visual speaker diarisation of real-world meetings. arXiv preprint arXiv:1906.10042  (2019)

\bibitem{DBLP:conf/interspeech/ChungNZ18}
Chung, J.S., Nagrani, A., Zisserman, A.: Voxceleb2: Deep speaker recognition. In: {INTERSPEECH}. pp. 1086--1090. {ISCA} (2018)

\bibitem{DBLP:conf/cvpr/DengGXZ19}
Deng, J., Guo, J., Xue, N., Zafeiriou, S.: Arcface: Additive angular margin loss for deep face recognition. In: {CVPR}. pp. 4690--4699. Computer Vision Foundation / {IEEE} (2019)

\bibitem{DBLP:conf/interspeech/DesplanquesTD20}
Desplanques, B., Thienpondt, J., Demuynck, K.: {ECAPA-TDNN:} emphasized channel attention, propagation and aggregation in {TDNN} based speaker verification. In: {INTERSPEECH}. pp. 3830--3834. {ISCA} (2020)

\bibitem{DBLP:conf/naacl/DevlinCLT19}
Devlin, J., Chang, M., Lee, K., Toutanova, K.: {BERT:} pre-training of deep bidirectional transformers for language understanding. In: {NAACL-HLT} {(1)}. pp. 4171--4186. Association for Computational Linguistics (2019)

\bibitem{DBLP:conf/icassp/FanKLLCCZZCW20}
Fan, Y., Kang, J.W., Li, L.T., Li, K.C., Chen, H.L., Cheng, S.T., Zhang, P.Y., Zhou, Z.Y., Cai, Y.Q., Wang, D.: Cn-celeb: {A} challenging chinese speaker recognition dataset. In: {ICASSP}. pp. 7604--7608. {IEEE} (2020)

\bibitem{gebru2017audio}
Gebru, I.D., Ba, S., Li, X., Horaud, R.: Audio-visual speaker diarization based on spatiotemporal bayesian fusion. IEEE transactions on pattern analysis and machine intelligence  \textbf{40}(5),  1086--1099 (2017)

\bibitem{DBLP:conf/cvpr/HeZRS16}
He, K., Zhang, X., Ren, S., Sun, J.: Deep residual learning for image recognition. In: {CVPR}. pp. 770--778. {IEEE} Computer Society (2016)

\bibitem{he2022}
He, M.k., Du, J., Lee, C.H.: End-to-end audio-visual neural speaker diarization. In: Proc. Interspeech 2022. pp. 1461--1465 (2022)

\bibitem{DBLP:journals/neco/HochreiterS97}
Hochreiter, S., Schmidhuber, J.: Long short-term memory. Neural Comput.  \textbf{9}(8),  1735--1780 (1997)

\bibitem{DBLP:journals/taslp/HsuBTLSM21}
Hsu, W., Bolte, B., Tsai, Y.H., Lakhotia, K., Salakhutdinov, R., Mohamed, A.: Hubert: Self-supervised speech representation learning by masked prediction of hidden units. {IEEE} {ACM} Trans. Audio Speech Lang. Process.  \textbf{29},  3451--3460 (2021)

\bibitem{DBLP:conf/mlmi/IstrateFMBB05}
Istrate, D., Fredouille, C., Meignier, S., Besacier, L., Bonastre, J.: {NIST} rt'05s evaluation: Pre-processing techniques and speaker diarization on multiple microphone meetings. In: {MLMI}. Lecture Notes in Computer Science, vol.~3869, pp. 428--439. Springer (2005)

\bibitem{liu2023multi}
Liu, T., Chen, Z., Qian, Y., Yu, K.: Multi-speaker end-to-end multi-modal speaker diarization system for the misp 2022 challenge. In: ICASSP 2023-2023 IEEE International Conference on Acoustics, Speech and Signal Processing (ICASSP). pp.~1--2. IEEE (2023)

\bibitem{DBLP:journals/taslp/MiroBEFFV12}
Mir{\'{o}}, X.A., Bozonnet, S., Evans, N.W.D., Fredouille, C., Friedland, G., Vinyals, O.: Speaker diarization: {A} review of recent research. {IEEE} Trans. Speech Audio Process.  \textbf{20}(2),  356--370 (2012)

\bibitem{DBLP:conf/interspeech/NagraniCZ17}
Nagrani, A., Chung, J.S., Zisserman, A.: Voxceleb: {A} large-scale speaker identification dataset. In: {INTERSPEECH}. pp. 2616--2620. {ISCA} (2017)

\bibitem{DBLP:conf/icassp/PanayotovCPK15}
Panayotov, V., Chen, G., Povey, D., Khudanpur, S.: Librispeech: An {ASR} corpus based on public domain audio books. In: {ICASSP}. pp. 5206--5210. {IEEE} (2015)

\bibitem{sarafianos2016audio}
Sarafianos, N., Giannakopoulos, T., Petridis, S.: Audio-visual speaker diarization using fisher linear semi-discriminant analysis. Multimedia Tools and Applications  \textbf{75},  115--130 (2016)

\bibitem{sharma2022using}
Sharma, R., Narayanan, S.: Using active speaker faces for diarization in tv shows. arXiv preprint arXiv:2203.15961  (2022)

\bibitem{tao2021someone}
Tao, R., Pan, Z., Das, R.K., Qian, X., Shou, M.Z., Li, H.: Is someone speaking? exploring long-term temporal features for audio-visual active speaker detection. In: Proceedings of the 29th ACM International Conference on Multimedia. pp. 3927--3935 (2021)

\bibitem{DBLP:conf/nips/VaswaniSPUJGKP17}
Vaswani, A., Shazeer, N., Parmar, N., Uszkoreit, J., Jones, L., Gomez, A.N., Kaiser, L., Polosukhin, I.: Attention is all you need. In: {NIPS}. pp. 5998--6008 (2017)

\bibitem{watanabe2020chime}
Watanabe, S., Mandel, M., et~al., B.J.: Chime-6 challenge: Tackling multispeaker speech recognition for unsegmented recordings. Interspeech  (2020)

\bibitem{wuerkaixi2022dyvise}
Wuerkaixi, A., Yan, K., Zhang, Y., Duan, Z., Zhang, C.: Dyvise: Dynamic vision-guided speaker embedding for audio-visual speaker diarization. In: 2022 IEEE 24th International Workshop on Multimedia Signal Processing (MMSP). pp.~1--6. IEEE (2022)

\bibitem{zhang2022npu}
Zhang, L., Li, Y., Wang, N., Liu, J., Xie, L.: Npu-hc speaker verification system for far-field speaker verification challenge 2022. INTERSPEECH (2022)

\bibitem{gulati2020conformer}
Zhang, Y., Lv, Z., et~al., W.H.: Mfa-conformer: Multi-scale feature aggregation conformer for automatic speaker verification. Interspeech  (2022)

\bibitem{zhou2022audio}
Zhou, H., Du, J., Zou, G., Nian, Z., Lee, C.H., Siniscalchi, S.M., Watanabe, S., Scharenborg, O., Chen, J.: Audio-visual wake word spotting in misp2021 challenge: Dataset release and deep analysis. In: Proceedings of the Annual Conference of the International Speech Communication Association, INTERSPEECH. vol.~2022 (2022)

\end{thebibliography}
\end{document}